\documentclass[%
 reprint,
superscriptaddress,
 amsmath,amssymb,
 aps,
]{revtex4-2}
\usepackage[compat=1.1.0]{tikz-feynman}
\usepackage{relsize}
\usepackage{slashed}
\usepackage{float}
\usepackage{array}
\usepackage{ulem}
\usepackage{graphicx}
\usepackage{dcolumn}
\usepackage{bm}

\usepackage{xcolor}

\begin{document}


\title{Type-3/2 Seesaw Mechanism}

\author{Durmu{\c s}  Demir}
\affiliation{%
Sabanc{\i} University, Faculty of Engineering and Natural Sciences, 34956 Tuzla {\.I}stanbul, Turkey
}%

\author{Canan Karahan}
\altaffiliation{\texttt{ckarahan@itu.edu.tr}}
\affiliation{%
 Physics Engineering Department,
{\.I}stanbul Technical University, 34469 Maslak {\.I}stanbul, Turkey
}%
\author{Ozan Sarg{\i}n}
\affiliation{%
{\.I}zmir Institute of Technology, Department of Physics, 35430, {\.I}zmir, Turkey
}

\date{\today}

\begin{abstract}
Type-I seesaw provides a natural explanation for the tiny neutrino masses. The right-handed neutrino masses it requires are, however, too large to keep the Higgs boson mass at its measured value. Here we show that vector-spinors, singlet leptons like the right-handed neutrinos, generate the tiny neutrino masses naturally
by the exchange of its spin-1/2 and spin-3/2 components. This one-step seesaw mechanism, which we call Type-3/2 seesaw, keeps the Higgs boson mass unchanged at one loop, and gives cause therefore to no fine-tuning problem. If the on-shell vector-spinor is a pure spin-3/2 particle then it becomes a potential candidate for hidden dark matter diluting due only to the expansion of the Universe. The  Type-3/2  seesaw provides a natural framework for the neutrino, Higgs boson and dark matter sectors, with overall agreement with current experiments and observations.   
\end{abstract}

\maketitle


{\emph{Introduction}\textemdash} Neutrino oscillations \cite{solar,atmospheric} are a proof that the active neutrinos are massive.  Neutrino mass, whose nature is still an open question, is a proof that there is new physics beyond the Standard Model (SM) \cite{minkowski,gell-mann,yanagida,goran,numas-np,pgl}.\\

Even though neutrino physics has come of age in the past two decades \cite{review-nu-mass2}, there is still no telling if neutrinos are Dirac ($\nu \neq \nu^c$) or  Majorana  ($\nu = \nu^c$) fermions  \cite{nonu2beta,Dolinski:2019nrj,Czakon:1999ed,pgl}. The Dirac masses \cite{Demir:2007dt,Bonilla:2016zef,Wang:2016lve,Wang:2017mcy,Yao:2018ekp,Calle:2018ovc,Saad:2019bqf,Jana:2019mez} conserve lepton number. The Dirac neutrinos acquire their masses via the electroweak breaking as all the other fermions, albeit with an unnaturally small Yukawa coupling. The Majorana masses, on the other hand, break lepton number, and arise via the electroweak breaking, with  naturally heavy SM-singlet right-handed neutrinos \cite{minkowski,gell-mann,yanagida,goran}.\\ 

The new physics that generates the tiny neutrino masses has been modeled variously by invoking various fields, symmetries and scales \cite{review-nu-mass4,review-nu-mass3,review-nu-mass2}. The Type-I seesaw mechanism, the first and the foremost of all \cite{gell-mann}, leads to small active neutrino masses \cite{gell-mann} via  the dimension-5 Weinberg operator \cite{wein-dim5} induced by the heavy right-handed neutrino mediation. There are also Type-II \cite{seesaw-II-a,seesaw-II-b,seesaw-II-c} and Type-III \cite{seesaw-III} seesaw mechanisms, which are mediated, respectively, by the  triplet scalars and triplet fermions. In addition to these tree-level mechanisms, various models \cite{Hou:1994et,Bamba:2008jq,Ma:2008cu,Ma:2009gu,Babu:2011vb,Hehn:2012kz,Ahriche:2014oda,Nomura:2016run,Ahriche:2017iar,Nomura:2020azp,Cai:2017jrq,Klein:2019iws} have been constructed to generate the small neutrino masses radiatively.\\

In this Letter we further the Type-I seesaw. We do this by replacing the right-handed neutrino with a vector-spinor field \cite{rs} involving both spin-3/2 and spin-1/2 components when it is off-shell and only a spin-3/2 component when it is on-shell.  We call the resulting neutrino mass generation mechanism  ``Type-3/2 seesaw mechanism". As will be shown in the sequel, this new mediator leads to important new results for the neutrino, Higgs and dark matter sectors.\\

Below, we first summarize the Type-I seesaw, and discuss its  implications for leptogenesis, dark matter and Higgs boson mass. Next, we  turn to the vector-spinor and construct the Type-3/2 seesaw, with a detailed discussion of its implications for leptogenesis, dark matter and the Higgs boson mass. Finally, we conclude the Letter by contrasting the salient features and implications of the Type-I and Type-3/2 seesaw mechanisms.\\

\begin{figure}[tpb]
    \centering
        \begin{tikzpicture}[baseline={(current bounding box.center)}] 
            \begin{feynman}
                 \vertex (a) ;
                 \vertex [above  =of a] (b) {\(\langle H \rangle\)};
                 \vertex [ left=of a] (d) {\(\nu_L\)};
                 \vertex [      right=of a] (f) ;
                 \vertex [      right=of f] (g) ;
                 \vertex [above =of g] (c) {\(\langle H \rangle\)};
                 \vertex [right=of g] (e) {\(\nu_L\)};
                 \diagram* {
                     (b) -- [scalar] (a) -- [anti fermion] (d),
                     (a) -- [red, fermion, edge label'=\(N\), insertion={[style=black, size=5pt]0.99999}] (f),
                     (f)-- [red, anti fermion, edge label'=\(N\)] (g),
                     (c) -- [scalar] (g) -- [anti fermion] (e),
                     };
                 \vertex [below=0.2cm of a] {\(y_N\)};
                 \vertex [above=0.2cm of f] {\(M_N\)};
                 \vertex [below=0.2cm of g] {\(y_N\)};
            \end{feynman}
       \end{tikzpicture}
    \caption{Type-I seesaw: Heavy right-handed neutrinos $N$ lead to the neutrino Majorana masses $m_\nu$.}
    \label{fig:seesaw_I}
\end{figure}
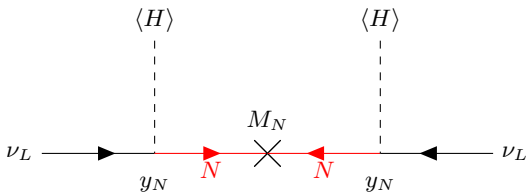

{\emph{Type-I Seesaw}\textemdash} The right-handed neutrino $N$, an SM-singlet spin-1/2 neutral  fermion \cite{minkowski,gell-mann}, gives a simple model of the tiny neutrino masses. It couples to the active neutrinos $\nu_L$ through the left-handed lepton doublet  $L=(\nu_L,\ell_L)$ as  
\begin{eqnarray}
\label{eqn:type1int}
    y_{N}\overline{L}HN + \frac{M_N}{2} \overline{N} N + {\rm H.C.} 
\end{eqnarray}
so that an active neutrino acquires the Majorana mass 
\begin{eqnarray}
    m_\nu = \frac{y_N^2 \langle H \rangle^2}{2 M_N}
    \label{mnu-RH}
\end{eqnarray}
via the Feynman diagram in Fig. \ref{fig:seesaw_I}. This induced mass, resulting from the Weinberg operator \cite{wein-dim5}, agrees with the experimental data  \cite{solar,atmospheric} for right-handed neutrino masses $M_N\approx 10^{14}\ {\rm GeV}$, Higgs vacuum expectation value 
$\langle H \rangle \approx 246\ {\rm GeV}$, and Yukawa coupling $y_N\approx {\mathcal{O}}(1)$. This dynamical mechanism, the Type-I seesaw \cite{minkowski,gell-mann,yanagida,goran}, generates neutrino Majorana masses $m_\nu$ naturally  with naturally heavy SM-singlet right-handed neutrinos \cite{numas-np}. Neutrino mixings \cite{review-nu-mass3,review-nu-mass2} are realized with two or more right-handed neutrinos. \\

\begin{figure}[htbp!]
  \centering
  \includegraphics{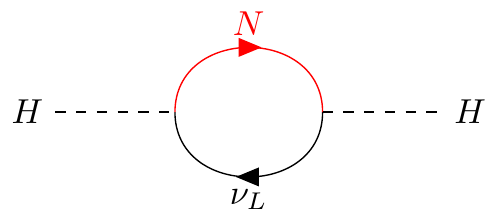}
    \caption{The active neutrino/right-handed neutrino loop that gives cause to the Higgs boson mass shift in (\ref{mh-shift}).}
    \label{fig:nuR_nuL_loop}
\end{figure}

The interactions in (\ref{eqn:type1int}) give cause for not only the neutrino masses as in (\ref{mnu-RH}) but also the Higgs  mass shift \cite{deltMH-MN}
\begin{eqnarray}
(\delta m_h^2)_N = \frac{m_\nu M_N^3}{ 2 \pi^2 \langle H \rangle^2} \log\frac{Q}{M_N} 
\label{mh-shift}
\end{eqnarray}
at the renormalization scale $Q\gtrsim M_N$ via the Feynman diagram in Fig. \ref{fig:nuR_nuL_loop}. This mass correction, evaluated in dimensional regularization  in which quadratic (and quartic) corrections all vanish identically \cite{dim-reg},  exceeds the Higgs boson mass itself unless $M_N \lesssim 10^{7}\ {\rm GeV}$  \cite{deltMH-MN1, deltMH-MN2,Bambhaniya:2016rbb}. This bound is in clear contradiction with the value $M_N \approx 10^{14}\ {\rm GeV}$ required by the neutrino masses. This contradiction shows that the right-handed neutrinos generate the active neutrino masses at the expense of an immense fine-tuning of the model parameters entering the Higgs boson mass \cite{Casas:2004gh}. This is a serious naturalness problem because the logarithmic correction (\ref{mh-shift}) survives to impede the Type-I seesaw  \cite{deltMH-MN2} even in the supersymmetry  \cite{Baer}. \\

The right-handed neutrinos can decay and annihilate \cite{rhn} via their interactions with the SM fields in (\ref{eqn:type1int}), and can facilitate therefore, for instance, the leptogenesis \cite{leptogenesis,leptogenesis2}.  Thermal leptogenesis  requires the lightest right-handed neutrino $N_1$ to have  a mass $M_{N_1} \gtrsim 2 \times 10^{9}\ {\rm GeV}$ in order to produce the requisite  asymmetry in the lepton sector. To be able to produce such a massive $N_1$ thermally, the reheat temperature $T_{rh}$ after the inflation must have a value $T_{rh} > M_{N_1}$  \cite{Buchmuller3,Buchmuller4,Buchmuller5}. It is clear that this leptogenesis value of $M_{N_1}$, too,  is in contradiction with the bound  $M_{N_1} \lesssim 10^{7}\ {\rm GeV}$ imposed by the Higgs boson mass correction in (\ref{mh-shift}). Even though it is not possible to suppress all the radiative corrections, sometimes gravity-mediated softly broken supersymmetry is incorporated into the thermal leptogenesis  to reduce the quadratic corrections from heavy right-handed neutrinos to the milder logarithmic ones.  This attempt leads, however, to the well-known gravitino problem \cite{Khlopov,Ellis}. It turns out that for gravitinos of masses below a few $\ {\rm TeV}$ the reheat temperature of the Universe should not exceed  $10^{5}\ {\rm GeV}$ \cite{Kawasaki,Cyburt}.\\

The Type-I seesaw does not offer a unique dark matter candidate though, as an analogous low-energy extension, one can consider incorporating the sterile neutrinos into the setup \cite{Boyarsky:2018tvu}.

{\emph{Type-3/2 Seesaw}\textemdash} Having discussed the Type-I seesaw and its physics implications, we now develop a new approach in which we envision the right-handed neutrino as spin-1/2 component of an SM-singlet, neutral vector-spinor $\psi_\mu$ of mass $M_\psi$. This Rarita-Schwinger field \cite{rs} decomposes as $\left[(1,\frac{1}{2})\oplus (0,\frac{1}{2})\right] \oplus \left[(\frac{1}{2},1)\oplus (\frac{1}{2},0)\right]$ under the Lorentz group. Namely, it involves spin-3/2 and spin-1/2 components. As was already pointed out in \cite{hidden}, it directly couples to the Higgs and lepton doublets via the Lagrangian
\begin{eqnarray}
\label{int1}
y_\psi \overline{L} H \gamma^{\mu}\psi_{\mu} + y_\psi \overline{\psi}_{\mu}  \gamma^{\mu} H^{\dagger} L  + \overline{\psi}_{\mu} \Lambda^{\mu \nu}\psi_\nu 
\end{eqnarray}
in which the kinetic  operator \cite{rs}
\begin{eqnarray}
\label{oper}
\Lambda^{\mu \nu} &=& \eta^{\mu \nu} (\slashed{p} - M_{\psi})  \nonumber\\ &-& (\gamma^{\mu}p^{\nu}+p^{\mu}\gamma^{\nu}) + 
\gamma^{\mu}\slashed{p}\gamma^{\nu} + M_{\psi} \gamma^\mu \gamma^\nu
\end{eqnarray}
differs from the Dirac operator by the terms in the second line. The Lagrangian (\ref{int1}) leads to the equation of motion
\begin{eqnarray}\label{eqnofmotion}
\Lambda_{\mu \nu} \psi^\nu = y_{\psi} \gamma_{\mu} H^{\dagger} L 
 \end{eqnarray}
in agreement with the vector-spinor description in \cite{rs2,rs3}.  For a free field, that is, for an on-shell vector-spinor $\psi^\mu_{(free)}(p)$ satisfying $p^2=M_{\psi}^2$ one gets the equation of motion $\Lambda_{\mu\nu}\psi_{(free)}^\nu=0$. This homogeneous equation can be consistently split into three distinct parts 
\begin{eqnarray}
(\slashed{p}-M_{\psi})\psi^{\mu}_{(free)} &=& 0,\label{eom1} \\
\gamma_{\mu}\psi^{\mu}_{(free)}&=&0,\label{eom2} \\ 
p_{\mu}\psi^{\mu}_{(free)}&=&0\label{eom3}
\end{eqnarray}
as follows from the kinetic structure in (\ref{oper}) as  particular choices for its individual terms. Here the point is that the equations (\ref{eom2}) and (\ref{eom3}) eliminate the spin-1/2 component of the vector-spinor $\psi^\mu_{(free)}$ to yield an on-shell pure spin-3/2 field. 

For an off-shell vector-spinor $\psi^\mu(p)$ satisfying $p^2\neq M_{\psi}^2$ the description is given by its propagator $S^{\mu\nu}$ \cite{rs2,rs3}
\begin{eqnarray}
\label{prop}
S^{\mu\nu} &=& -\frac{i \eta^{\mu\nu}}{\slashed{p}-M_{\psi}} \nonumber\\
&+&\frac{i}{\slashed{p}-M_{\psi}}\!\left(\! \frac{\gamma^\mu\gamma^\nu}{3} + \frac{(\gamma^\mu p^\nu - p^\mu \gamma^\nu)}{3 M_{\psi}} + \frac{2 p^\mu p^\nu}{ 3 M_{\psi}^2}\!\right)
\end{eqnarray}
whose second line stands for deviation from the spin-1/2 propagator in parallel with (\ref{oper}). This propagator is nothing but the inverse of the kinetic structure in (\ref{oper}) namely $S^{\mu\alpha} \Lambda_{\alpha \nu} = \delta^\mu_\nu$. This propagator holds in the entire momentum and spin space \cite{yeni}. In other words, this vector-spinor propagator involves propagations of both spin-3/2 and spin-1/2 components. It certainly relates $\psi_\mu$ to its source in (\ref{eqnofmotion}) but it does not satisfy the individual motion equations in (\ref{eom1}), (\ref{eom2}) and (\ref{eom3}). 

It proves useful to contrast the vector-spinor above with the  gravitino  \cite{neu,strumia} -- the gauge field of supergravity which acquires mass by swallowing the goldstino field \cite{sugra}. The gravitino is a spin-3/2 field and therefore obeys the equations (\ref{eom1}), (\ref{eom2}) and (\ref{eom3}) when it is on-shell and off-shell. The gravitino propagator is constructed in \cite{neu} by imposing (\ref{eom1}), (\ref{eom2}) and (\ref{eom3}) (with  (\ref{eom2}) and (\ref{eom3}) being kind of gauge conditions) so that as a propagator it satisfies (\ref{eom1}), (\ref{eom2}) and (\ref{eom3}). 

The vector-spinor $\psi_\mu$ is different than the gravitino. Its propagator in (\ref{prop}) propagates both spin-3/2 and spin-1/2 components, and does not obey the motion equations (\ref{eom1}), (\ref{eom2}) and (\ref{eom3}). It can be reduced to a spin-3/2 particle when it is on-shell, that is, when it obeys the individual motion equations (\ref{eom1}), (\ref{eom2}) and (\ref{eom3}). Its electric neutrality ensures that there arises no problem with local causality 
 \cite{local_causal_1,local_causal_2}.

\begin{figure}[tpb]
    \centering
        \begin{tikzpicture}[baseline={(current bounding box.center)}] 
            \begin{feynman}
                 \vertex (a) ;
                 \vertex [above  =of a] (b) {\(\langle H \rangle\)};
                 \vertex [ left=of a] (d) {\(\nu_L\)};
                 \vertex [      right=of a] (f) ;
                 \vertex [      right=of f] (g) ;
                 \vertex [above =of g] (c) {\(\langle H \rangle\)};
                 \vertex [right=of g] (e) {\(\nu_L\)};
                 \diagram* {
                     (b) -- [scalar] (a) -- [anti fermion] (d),
                     (a) -- [blue, fermion, boson, edge label'=\(\psi\), insertion={[style=black, size=5pt]0.99999}] (f),
                     (f) -- [blue, anti fermion, boson, edge label'=\(\psi\)] (g),
                     (c) -- [scalar] (g) -- [anti fermion] (e),
                     };
                 \vertex [below=0.2cm of a] {\(y_{\psi}\)};
                 \vertex [above=0.2cm of f] {\(M_{\psi}\)};
                 \vertex [below=0.2cm of g] {\(y_{\psi}\)};
            \end{feynman}
       \end{tikzpicture}
    \caption{Type-3/2 seesaw: Heavy vector-spinor neutrinos $\psi_\mu$ lead to the neutrino Majorana masses $m_\nu$.}
    \label{fig:seesaw_32}
\end{figure}
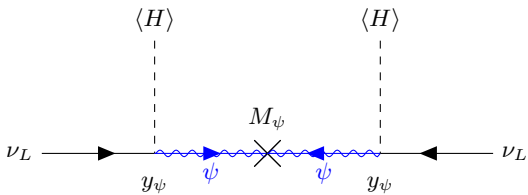

In similarity to the right-handed neutrinos, the $\psi_\mu$ couplings in (\ref{int1}) lead to the active neutrino masses
\begin{eqnarray}
    m_\nu = \frac{2 y_{\psi}^2 \langle H \rangle^2}{9 M_{\psi}}
\label{mnu-RS}
\end{eqnarray}
via the Feynman diagram in Fig. \ref{fig:seesaw_32} with the exchange of spin-3/2 and spin-1/2 components. This result, derived by using the $\psi_\mu$ propagator in (\ref{prop}), agrees with the experimental data \cite{solar,atmospheric} for $M_{\psi}\approx 10^{14}\ {\rm GeV}$ and $y_{\psi}\approx {\mathcal{O}}(1)$. This new mechanism, which is what we term as Type-3/2 seesaw, generates the neutrino Majorana masses naturally with a naturally heavy SM-singlet vector-spinor $\psi_\mu$. Neutrino mixings \cite{review-nu-mass3,review-nu-mass2} can be realized with two or more $\psi_\mu$ fields. \\

If the vector-spinor $\psi_\mu$ obeys the three equations of motion  (\ref{eom1}), (\ref{eom2}) and (\ref{eom3}) then it becomes a spin-3/2 field when it is on-shell and in this case, in contrast to the right-handed neutrinos, it can neither decay nor annihilate despite its couplings in (\ref{int1}) to the SM fields. The reason is that in such processes $\psi_\mu$ is on its mass-shell as an isolated physical particle and for an on-shell $\psi_\mu$ namely for a $\psi_\mu$ satisfying (\ref{eom1}), (\ref{eom2}) and (\ref{eom3}) the $H$-$L$-$\psi_\mu$ vertex vanishes identically  (\cite{rs2,rs3}). (This vertex vanishes for both on-shell and off-shell gravitino \cite{neu}.)  This means that the scattering processes with on-shell $\psi_\mu$ (its decays, annihilations and productions) all vanish. This on-shell nullity of $\psi_\mu$ has three important implications. Firstly, $\psi_\mu$, unlike the right-handed neutrinos, cannot facilitate leptogenesis simply because there exists no decay channel to transfer lepton number in $\psi_\mu$ to active neutrinos and charged leptons \cite{ozan}. Needless to say, baryogenesis can occur via some other mechanism such as Affleck-Dine \cite{aff-din} baryogenesis, and this sets the theory exempt from the gravitino problem \cite{Weinberg2} in gravity-mediated softly-broken supersymmetry \cite{sugra}. \\

Secondly, if on-shell $\psi_\mu$ is a spin-3/2 particle obeying (\ref{eom1}), (\ref{eom2}) and (\ref{eom3})  then it is an everlasting particle. It was around, is around and will be around to participate in certain processes in a hidden or invisible way \cite{hidden}. Its density falls with the volume of the Universe as it dilutes due only to the expansion of the Universe. It can therefore form  dark matter \cite{cdm} if it leads to flat rotation curves, structure formation and other relevant phenomena. The setup in (\ref{int1}) can be modified to obtain a detectable and more conventional dark matter candidate. One possibility is to invoke higher-dimension operators \cite{cdm-higher-dim} but the $H$-$L$-$\psi_\mu$  coupling in (\ref{int1}) must still be taken into account when $\psi_\mu$ is off-shell.\\

\begin{figure}[htbp!]
  \centering
  \includegraphics{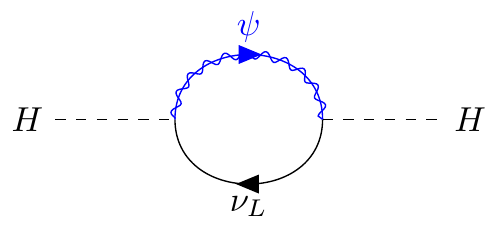}
    \caption{The vector-spinor/active neutrino loop that does not lead to any Higgs boson mass shift.}
    \label{fig:psi_nuL_loop}
\end{figure}

Thirdly, when it is off-shell $\psi_\mu$ reveals itself via its spin-3/2 and spin-1/2 components \cite{hidden} by inducing the neutrino Majorana masses as in (\ref{mnu-RS}), 
altering certain SM scattering amplitudes such as $h h \rightarrow {\overline{\nu_L}} \nu_L$ \cite{hidden},  and facilitating loop corrections to certain SM masses and couplings. Its loop with the active neutrinos, depicted in Fig. \ref{fig:psi_nuL_loop}, is expected to shift the Higgs boson mass as in (\ref{mh-shift}) but, to one's surprise, it gives actually zero contribution
\begin{eqnarray}
(\delta m_h^2)_{\psi_\mu} = 0
\label{mh-shift2}
\end{eqnarray}
at one loop. This follows from the fact that the Feynman diagram in Fig. \ref{fig:psi_nuL_loop} evaluates to zero
\begin{widetext}
\begin{eqnarray}
&& -i \frac{y_{\psi}^2}{2} \int \frac{d^4 p}{(2\pi)^4} \frac{1}{p^2 (p^2 - M_{\psi}^2)} {\rm Tr} \left[P_R \gamma_\alpha (\slashed{p} + M_{\psi})\left\{ \eta^{\alpha\beta} -\frac{\gamma^\alpha\gamma^\beta}{3} - \frac{2 p^\alpha p^\beta}{3 M_{\psi}^2} - \frac{\left(\gamma^\alpha p^\beta - p^\alpha \gamma^\beta\right)}{3 M_{\psi}}\right\} \gamma_\beta \slashed{p} P_R \right]\nonumber\\
&=& -i y_{\psi}^2 M_{\psi}^2 \lim_{\epsilon \rightarrow 0} Q^\epsilon \left(\frac{M_{\psi}^2}{4\pi}\right)^{2-\frac{\epsilon}{2}} \left[ \Gamma\left(-1+\frac{\epsilon}{2}\right)
-  \left(-2+\frac{\epsilon}{2}\right) \Gamma\left(-2+\frac{\epsilon}{2}\right)
\right] = 0
\end{eqnarray}
\end{widetext}
in confirmation of the nullity of $\delta^4(0)$ in dimensional regularization \cite{dim-reg}. This zero Higgs mass shift, parametrized by the $\psi_\mu$ momentum $p_\alpha$, right projector $P_R=(1+\gamma_5)/2$, and the renormalization scale $Q$, is evaluated in the same  regularization scheme as the Higgs mass shift in (\ref{mh-shift}). It ensures that  the active neutrinos and the Higgs boson can acquire their measured masses with no contradiction concerning the scale of $M_{3/2}$. The naturalness (fine-tuning) problem impeding the Type-I seesaw does simply not exist in the Type-3/2 seesaw, owing to (\ref{mh-shift2}). \\


{\emph{Conclusion}\textemdash} In this Letter we have shown that the SM-singlet vector-spinor \cite{rs} gives rise to a new one-step seesaw mechanism. This new mechanism, the Type-3/2 seesaw, exhibits physically important features not found in the Type-I seesaw. The salient features are:
\begin{itemize}
    \item Tiny neutrino Majorana masses arise naturally  (in agreement with the Type-I seesaw),
    
    \item Higgs boson acquires its mass without fine-tuning  (in disagreement with the Type-I seesaw),
    
    \item If the on-shell vector-spinor is a spin-3/2 field then baryogenesis is not sourced by leptogenesis (in disagreement with the Type-I seesaw), and this makes the model impervious to the gravitino problem in the  Type-I seesaw,  and finally
    
    \item If the on-shell vector-spinor is a spin-3/2 field  dark matter can exist as an everlasting field hidden in the SM spectrum  (in disagreement with Type-I seesaw). 
    
\end{itemize}
These points are tabulated in Table \ref{table:1} in a comparative fashion. In view of them,  one can conclude that the Type-3/2 seesaw proposed in this Letter can open up a novel approach to neutrino and dark sector phenomenology, with its inherent naturalness in both the neutrino and Higgs sectors. The minimal structure considered in this Letter can be extended to make the spin-3/2 dark matter detectable. This can be done in various ways such as  using   higher-dimension operators as in \cite{cdm-higher-dim, cdm-higher-dim2} or adding new fields, such as a scalar field coupling to ${\overline{\psi_\mu}} \psi^\mu$ \cite{canan-ozan}.

\begin{table}[h]
\centering
\caption{A comparative summary of the implications of the Type-I and Type-3/2 seesaw mechanisms. The label $^{(a)}$ is a warning  that the Type-I seesaw does not bring up any distinctive  dark matter candidate. The label $^{(b)}$, on the other hand, is a reminder that baryogenesis must occur  via mechanisms other than leptogenesis.}
\begin{tabular}[t]{lp{2.5cm}p{2.5cm}}
\hline
  & Type-I Seesaw  & Type-$3/2$ Seesaw \\ [0.5ex] 
\hline\hline
 $\nu$-oscillations & Yes & Yes  \\ 
 Stable Higgs mass & No & Yes  \\
 Dark matter & No$^{(a)}$ & Yes  \\
 Leptogenesis & Yes & No$^{(b)}$  \\
 \hline\hline
\end{tabular}
\label{table:1}
\end{table}


\begin{acknowledgments}
    The work of CK is supported by {\.I}T{\"U} BAP grant TAB-2020-42312. 
\end{acknowledgments}

\end{document}